\documentclass[]{article}

\usepackage{hyperref}
\usepackage{t1enc}
\usepackage{cite}
\usepackage{amsmath,empheq}
\usepackage[printonlyused]{acronym}
\usepackage{algorithm}
\usepackage{algorithmicx}
\usepackage[noend]{algpseudocode}
\usepackage{graphicx}
\usepackage{capt-of,etoolbox}
\usepackage{authblk}

\title{\Large A Physically Plausible Model for Rendering Highly Scattering Fluorescent Participating Media}

\author{Marwan Abdellah}
\author{Ahmet Bilgili}
\author{Stefan Eilemann}
\author{Henry Markram}
\author{Felix Sch\"urmann}

\affil{\small Blue Brain Project(BBP)\\ \'Ecole polytechnique f\'ed\'erale de Lausanne (EPFL)}
\date{} 
\begin{document}
    
\maketitle

\newcommand\myfigure{%
    \begin{center}
        \vspace*{-1cm}
    \includegraphics[width=.99\textwidth]{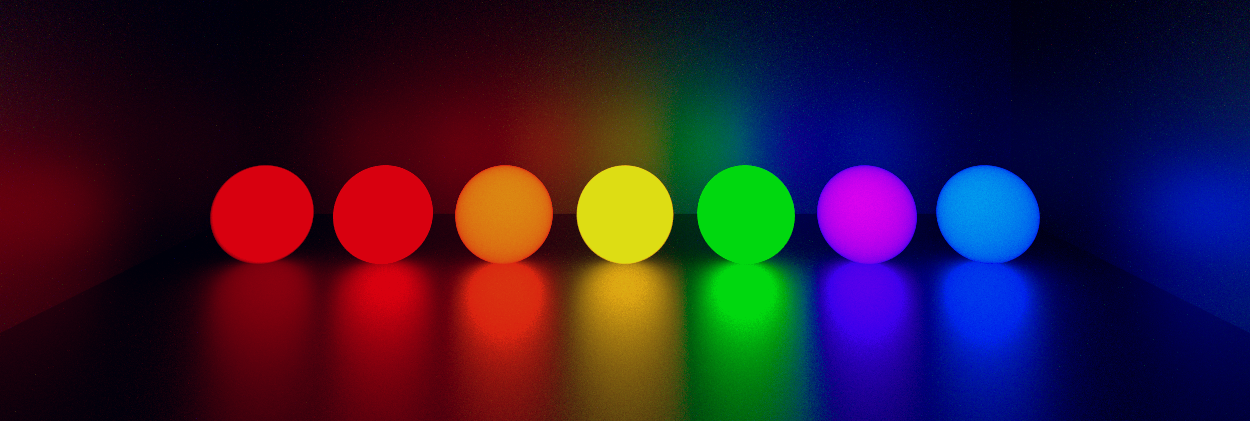}
    \captionof{figure}{Fluorescent spheres illuminating Cornell box. The spheres are filled with different fluorescent dyes (fluorophores) dissolved in highly scattering solution. The scene is illuminated with multiple wavelengths corresponding to the maximum excitation wavelength of each fluorophore. The emission spectra of the selected fluorophores cover the visible spectrum (300 and 800 nm), which seems apparent from the glossy reflection on the back wall of the Cornell box.}
    \label{fig:teaser-fluorescent-spheres}
\end{center}
}

\myfigure{}

\begin{abstract}
We present a novel extension of the path tracing algorithm that is capable of treating highly scattering participating media in the presence of fluorescent structures. The extension is based on the formulation of the full radiative transfer equation when solved on a per-wavelength-basis, resulting in accurate model and unbiased algorithm for rendering highly scattering fluorescent participating media. The model accounts for the intrinsic properties of fluorescent dyes including their absorption and emission spectra, molar absorptivity and quantum yield and also their concentration. Our algorithm is applied to render highly scattering isotropic fluorescent solutions under different illumination conditions. The spectral performance of the model is validated against emission spectra of different fluorescent dyes that are of significance in spectroscopy.
\end{abstract}

\section{Introduction}
\subsection{Motivation}
The history of computer graphics has witnessed extensive research efforts to solve the global illumination problem when the influence of participating media is taken into account. Various mathematical models and rendering algorithms have been proposed to address this problem and simulate light interaction with participating media on a physically-plausible basis~\cite{jensen1998efficient,cerezo2005survey}. 
Unbiased rendering methods, such as path tracing~\cite{raab2008unbiased} or Metropolis light transport~\cite{pauly2000metropolis}, are capable of obtaining accurate solutions for the radiative transfer equation (\acs{RTE}), achieving exceptional highly realistic synthetic images that are indistinguishable from real photographs. 
With few exceptions, the majority of rendering algorithms focused on handling absorption and scattering effects, and obviated physically-accurate models for rendering other phenomena such as self-emission~\cite{wilkie2011physically}, polarization~\cite{wilkie2001combined, wolff1990ray} and fluorescence.

Fluorescence is the process of absorption and re-emission of light at longer wavelengths. Many natural compounds present in our daily life exhibit fluorescence. They have several practical applications in various fields including spectroscopy, cell biology and microscopy. 

In general, rendering fluorescence has received little consideration in computer graphics literature and has been only investigated in few research studies for several reasons. 
Contrary to light scattering and absorption, fluorescence can be na\"ively faked with customized post-processing shaders, avoiding the effort required to model the underlying physics of the phenomenon itself~\cite{wilkie2001combined}. 
Public rendering engines have neglected fluorescence justifying its little practical value~\cite{pbrt2012}.
Moreover, solving the \acs{RTE} to estimate the radiance on a per-wavelength basis was extremely less efficient than combining multiple wavelengths at once relying on the tristimulus theory of light representation~\cite{raab2008unbiased}. Therefore, rendering fluorescent scenes was quite inapplicable in the past due to the non-existence of convenient spectral rendering engines that could efficiently handle wavelength-based lighting calculations such as PBRT~\cite{pbrt2012}, Mitsuba~\cite{mitsuba2010} or LuxRender~\cite{luxrender2013}.

During the recent years, rendering fluorescent participating media on a physically-plausible basis has turned out to be a necessity in certain fields where accurate prediction of the object appearance is a must.
This fact can be demonstrated by the following examples: 
\begin{enumerate}
    \item Simulation of underwater vision, which requires rendering the fluorescent pigments of the ocean waters~\cite{cerezo2003inelastic}.
    \item Simulation of optical sections in light sheet fluorescence microscopy~\cite{abdellah2015physically}.
    \item Simulation of the fluorescent structures in cell biology for in silico experimentation~\cite{sharpe2008silico}.
\end{enumerate}

In this paper, we present an extension of the path tracing algorithm based on accurate Monte Carlo model for rendering highly scattering fluorescent participating media in terms of their intrinsic spectral properties. 

\subsection{Contribution}
\begin{enumerate}
    \item Development of accurate mathematical model for rendering fluorescent dyes, taking into account their spectroscopic characteristic including absorption and emission spectra, molar absorptivity and quantum efficiency and also their concentration in the medium.  
    \item Extension of the standard path tracing algorithm to be capable of handling highly scattering participating media in the presence of fluorescent mixtures and applying the algorithm to render fluorescent solutions. 
    \item Quantitative validation of the spectral performance of the presented fluorescence model against emission spectra of several fluorescent dyes that have many significant applications in spectroscopy. 
\end{enumerate}

\section{Background and Related Work}
Accurate modeling of fluorescence requires decent knowledge of the basic physical principles of the phenomenon on both molecular and macroscopic levels. This is mandatory to understand how fluorescent materials can transfer the energy absorbed from one wavelength to another. In the following sections, we briefly review the physics of fluorescence, and then discuss the previous approaches for modeling the phenomenon and their limitations.

\subsection{Fluorescence Physics}
Fluorescence emission can occur when the molecules of certain chemical compounds (called \emph{fluorophores}) are excited due to light absorption, mechanical friction or chemical reaction~\cite{johnson2010molecular}. The majority of fluorescence phenomena in nature are caused by light absorption at certain wavelength and re-radiation at slightly longer one. The energy difference between the two wavelengths (\emph{Stokes shift}) is considered a fluorophore intrinsic property. The higher the Stokes shift, the better the fluorophore is. 

\begin{figure}[!h]
    \centering
    \mbox{} \hfill
    \includegraphics[width=0.90\columnwidth]{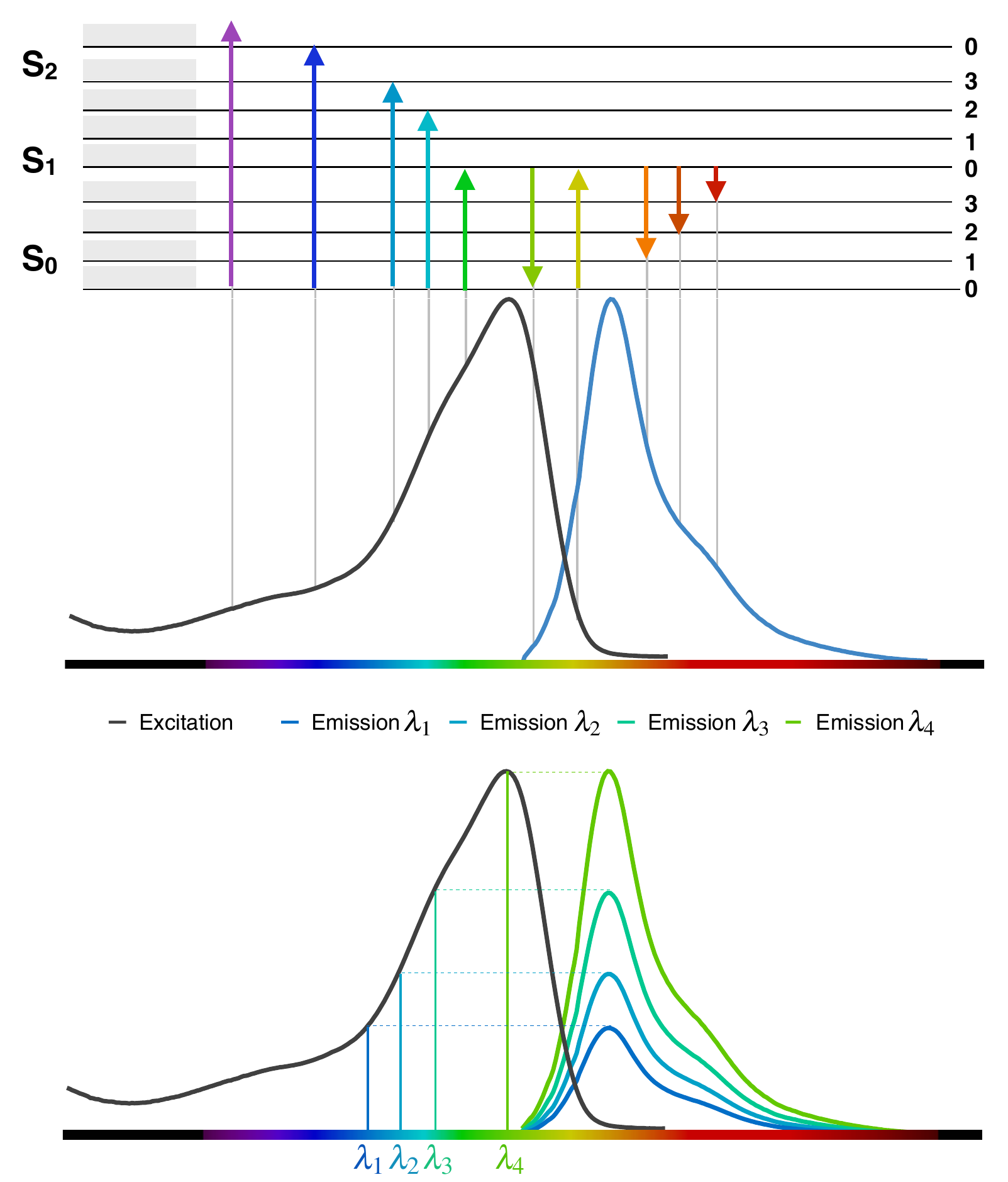}
    \hfill \mbox{}
    \caption{\label{fig:fluorescence-physics}
        (Top) The physics of fluorescence explained with a Jablonski diagram. 
        (Bottom) Excitation of a polyatomic fluorophore at four different wavelengths $\lambda_1$, $\lambda_2$, $\lambda_3$ and $\lambda_4$. Although the resulting emission spectra have the same profile, they have different intensities that correspond to the amplitude of the excitation spectrum at each illuminating wavelength . 
    }
\end{figure}

Each fluorophore has its distinct excitation and emission wavelengths that characterize its spectral behavior. Monoatomic compounds have discrete wavelengths, while polyatomic ones exhibit broad spectra~\cite{johnson2010molecular}. After photon absorption, the fluorophore is excited from the ground state to a higher electronic state. This excitation is followed by quick relaxation to the lowest vibrational level of the excited electronic state by internal conversion. Fluorescence emission occurs when the fluorophore returns to the electronic ground state. The transitions in the vibrational states in the excited electronic state are describe by transition probabilities, which accounts for the continuous profile of the emission spectrum. 
This is graphically explained in the Jablonski diagram shown in Figure~(\ref{fig:fluorescence-physics}). 
The emission and excitation wavelengths are independent, but the emission intensity is proportional to the amplitude of the excitation wavelength. Therefore, the excitation of a polyatomic fluorophore with monochromatic illumination will scale the the emission spectrum according to the amplitude of the excitation one at the exciting wavelength, see Figure~(\ref{fig:fluorescence-physics}) .  

The time difference between absorption and emission (\emph{fluorophore lifetime}) is in the range of few nanoseconds, and thus it can be ignored under stead state conditions~\cite{pharr2004physically}. The fluorescence emission is always assumed to be isotropic, thus, any fluorescence emission event will scatter with isotropic phase function. 


%

\subsection{Related Work: Fluorescence in Computer Graphics}
In 1995, Glassner~\cite{glassner1995principles} presented the first extension of the rendering equation to handle complex scenes including fluorescent materials. His work has been focused on adapting the scattering function in the rendering integral to account for fluorescence emission. He introduced the fluorescence efficiency term $f_{\zeta}(\lambda \leftarrow \lambda')$ to model the transfer of energy from one wavelength to another. Nevertheless, the mathematical formalism of his modeling work did not explicitly account for the distinct properties of fluorescent materials.
Further extensions to Glassner's model have been developed to treat the fluorescence as a volumetric \emph{inelastic} scattering effect that involves energy transfer when light scatters in participating media. 

The limitations of Glassner's work have been radically improved in following studies to handle the spectral behavior of the medium in terms of its characteristic parameters. 
Cerezo and Seron~\cite{cerezo2003inelastic, cerezo2004rendering} have developed a system for rendering non-homogeneous and anisotropic fluorescent participating media based on the discrete coordinates method. Their system was applied to study the effects of fluorescence in natural waters .
Gutierrez \emph{et al.} \cite{gutierrez2004inelastic, gutierrez2008visualizing} have presented another extension of the photon mapping algorithm based on the full \acs{RTE} and Fermat's law to handle multiple inelastic scattering and curved light paths in participating media with varying index of refraction. 
Although they were capable or rendering scattering fluorescent participating media, these studies were limited in two aspects. Their global illumination solutions were based on biased rendering algorithms (discrete ordinates and curved photon mapping). Moreover, they ignored the actual spectral properties of the fluorescent dyes and used approximated profiles for their excitation and emission spectra. 

Abdellah \emph{et al.}~\cite{abdellah2015computational,abdellah2015physically} have presented an unbiased solution for rendering pure fluorescent material that are normally characterized with low scattering properties taking into account the intrinsic spectroscopic characteristics of fluorescent dyes. Their presented framework was used for simulating imaging experiments in fluorescence microscopy, but it was only valid for rendering clarified brain models that have negligible scattering properties. 
Few other extensions to Glassner's model have been discussed to treat the fluorescence as a surface phenomenon using bispectral bidirectional reflectance
and reradiation distribution functions (or bispectral \acs{BRRDF}) and re-radiation matrices~\cite{glassner1995model, wilkie2001combined, wilkie2006reflectance, hullin2010acquisition, lam2013spectral}, but the discussion of these studies is out of scope.  

The work presented in this paper addresses the shortcomings of the previous approaches to provide a rendering algorithm for rendering highly scattering fluorescent participating media in terms of their spectroscopic properties in an unbiased manner. 

\section{Monte Carlo Fluorescence Modeling}
Numerous fluorescence Monte Carlo models have been proposed in the optics literature for simulating imaging experiments of scattering tissue, for example, standard fluorescence models, path-history-based decoupled models, and perturbation-based models. An extensive review of these models, their precision and their relative performance is given in~\cite{jiang2014evaluation, luo2015decoupled}. The models are based on the extension discussed by Welch \emph{et al.}~\cite{welch1997propagation} to account for (1) photon propagation in scattering media at excitation and emission wavelengths and (2) the excitation-to-emission conversion efficiency. The simulations use the photon tracing algorithm (particle model of light) to estimate the amount of fluorescence reaching the detector surface at specific filed of view. The exciting photons are normally sent from collimated or diffuse light sources to propagate in the medium until a probabilistic fluorescence emission happens. The emission photons do several bounces until being recorded by the detector or escaping the medium. 

This approach could be useful for optical imaging simulations under certain conditions, but it cannot be beneficial to computer graphics to render fluorescent scenes with merely participating media. The solution obtained relying on photon tracing models requires huge amount of samples to converge; since the majority of the photon bundles keep bouncing in the scene without contributing to the final radiance computed per pixel. Therefore rendering a meaningful fluorescent images that have converged solution will not be practically feasible. A partial solution that could be relatively practical would not be significant neither~\cite{pattanaik1993computation,cerezo2005survey}. 

\section{Fluorescence Path Tracing Algorithm}
The path tracing algorithm was introduced by Kajia~\cite{kajiya1986rendering} and extended later by Raab \emph{et al.}~\cite{raab2008unbiased} to consider the effects caused by participating media. 
This Monte Carlo algorithm obtains an unbiased solution to the light transport problem.
The paths are generated incrementally by sampling multiple scattering events starting at the camera and ending at the light source. 
Unlike weighted photon models~\cite{welch1997propagation,swartling2003accelerated}, absorption is implicitly handled in the transmittance computed between the scattering events relying on Beer Lambert law. 

Conventional path tracing algorithms solve the \acs{RTE} by computing tristimulus transport using the RGB color space. This approach is quite unfaithful to produce physically-plausible renderings when accurate prediction of object appearance is required~\cite{wilkie2014hero}. Meng \emph{et al.}~\cite{meng2015physically} introduced a precise mapping technique to improve the solution obtained with RGB-based path tracers. However, this mapping cannot be useful for representing the color -- or Stokes -- shift accompanying fluorescence emission. 
Advanced spectral path tracers have been presented to solve the \acs{RTE} on a per-wavelength basis, for example PBRT~\cite{pbrt2012} and Maxwell Render~\cite{render2010maxwell}. Their goal was focused on accurate rendering of objects characterized with complex spectral power densities (\acs{SPD}s), but they have not considered fluorescence. In the following part, we present our Monte Carlo fluorescence model and its implementation with adapted path tracing algorithm that can accurately solve the full \acs{RTE} when rendering scenes containing highly scattering fluorescent participating media. 

\subsection{Fluorescence Model}
Non fluorescent participating media are modeled in computer graphics as a collection of randomly positioned microscopic particles. 
The light interaction with those particles is assumed to be probabilistic. 
The scattering events along the light path are randomly sampled according to the optical properties of the medium.
Further assumptions are made when light interacts with fluorescent media. 
Pure fluorophores characterized with low scattering properties can be modeled with only fluorescent particles~\cite{abdellah2015physically}. 
When these fluorophores are diffused in highly scattering media, the light can interact with either fluorescent or non-fluorescent particles. 
An interaction with a fluorescent particle does not necessarily result in fluorescence emission, though, an emission event must be sampled at a fluorescent particle.
The scattering events in this case are sampled in terms of the concentration of the fluorophores diffused in the medium in addition to their absorption spectrum, molar absorptivity and quantum efficiency. 

\begin{figure}[!h]
    \centering
    \mbox{} \hfill
    \includegraphics[width=0.95\columnwidth]{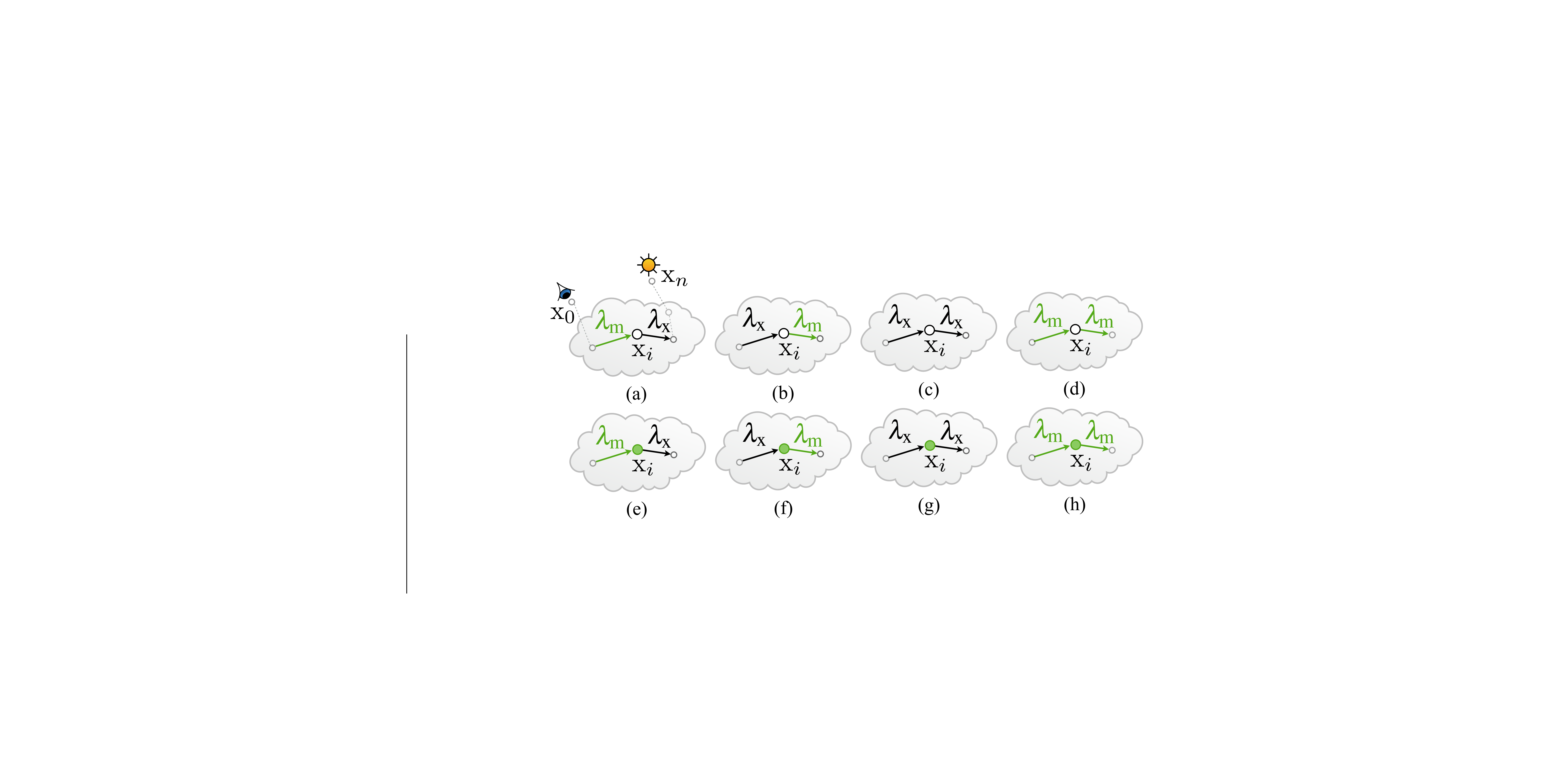}
    \hfill \mbox{}
    \caption{\label{fig:events-combinations}
        All the possible combination of events that might occur when a fluorescence scattering medium is sampled. 
    }
\end{figure}

In photon tracing, excitation photons are sent from the light source and the emission ones are received by the camera. The adjoint solution with path tracing requires sending emission photons from the camera and sampling the light source after the occurrence of fluorescence emission. 
As a consequence of neglecting secondary emission, each probabilistic path connecting the camera to the light source must consist of multiple elastic scattering events and only a single inelastic one. 
All the possible combination of events that can occur during the path generation in a fluorescent medium are shown in Figure~(\ref{fig:events-combinations}). The events in (a) and (b)  are not physically-plausible because a fluorescence emission requires a fluorescent particle to exist. The event in (f) is also invalid because an emission photon cannot excite the fluorophores to emit an excitation photon. The other events are probable, where (c), (d), (g) and (h) represent elastic scattering and (e) accounts for fluorescence emission. 

Unlike conventional path tracing, the light source is only sampled when a fluorescence emission event is realized. At this point, the excitation-to-emission conversion is evaluated to account for the amount of energy re-radiated at emission wavelength upon direct photon absorption from the light source at excitation wavelength.
Further events can be sampled on the path after this emission, but if another fluorescence emission happens, the event is ignored and the path is terminated. 
Elastic scattering events at excitation wavelength are also probable, however, their contribution is almost negligible after few bounces in the medium due to its high optical thickness.
Consequently, after the first emission event, the contribution of the path is added to pixel value and the path is terminated.    
If the emission photon keeps bouncing in the medium until escaping its boundaries before a fluorescence emission occurs, the path is ignored since it does not contribute to the radiance arriving to the pixel due to fluorescence emission.

\subsection{Path Integral Formulation}

Assuming steady-state conditions, if a non-monochromatic light source illuminates a scattering participating medium, the radiance $L$ reaching the camera film after a single scattering event within the extent of the medium is computed with the following formula 


\begin{equation} \label{eq:rte-sphere-form}
\begin{gathered}
L(\mathrm{x}_{0}, \omega) = \
\displaystyle \int_{\Omega_{4 \pi}} \, 
F(\mathrm{x}_{n}, \mathrm{x}_{i}, \mathrm{x}_{0}) \,
L_{i}(\mathrm{x}_{i}, \omega') \,
\mathrm{d}\omega'
\end{gathered}  
\end{equation} 

\begin{equation} \label{eq:rte-sphere-form}
\begin{gathered}
F(\mathrm{x}_{n}, \mathrm{x}_{i}, \mathrm{x}_{0}) = 
\sigma_s(\mathrm{x}_{i}) \, f_{\phi}(\mathrm{x}_{i}, \omega, \omega') 
\mathrm{d}\omega'
\end{gathered}  
\end{equation} 

where $\mathrm{x}_{0}$, $\mathrm{x}_{i}$ and $\mathrm{x}_{n}$ are randomly sampled points on the surface of the camera film, in the medium and on the emitting surface of the light source respectively, $\omega$ and $\omega'$ are the directions from $\mathrm{x}_{i}$ to $\mathrm{x}_{0}$ and $\mathrm{x}_{i}$ to $\mathrm{x}_{n}$, $L_{i}(\mathrm{x}_{i}, \omega')$ is the incident radiance arriving to the point $\mathrm{x}_{i}$ from the light source, $F$, $\sigma_s$ and $f_{\phi}$ are the scattering function, scattering coefficient and the phase function of the participating medium. For convenience, the integral over directions on the sphere ($\Omega_{4 \pi}$) in Equation~(\ref{eq:rte-sphere-form}) is rewritten as an integral over the surfaces and volumes in the scene to be 

\begin{equation} \label{eq:rte-three-point-form}
\begin{gathered}
L(\mathrm{x}, \omega) = \
\displaystyle \int_{A} \, 
F(\mathrm{x}_{n}, \mathrm{x}_{i}, \mathrm{x}_{0}) \, 
L_{r}(\mathrm{x}_{n}, \mathrm{x}_{i}) \,
V(\mathrm{x}_{n}, \mathrm{x}_{i}) \,
G(\mathrm{x}_{n}, \mathrm{x}) \,
\mathrm{d}A(\mathrm{x}_{n})
\end{gathered}  
\end{equation} 
\vspace*{-0.5cm}

where $V$ is the \emph{binary visibility function} that is equivalent to unity if the points $\mathrm{x}_{n}$ and $\mathrm{x}_{i}$ are mutually visible and zero otherwise, $G$ is the \emph{geometric term} that arises as a consequence of changing the integration variable where 

\begin{equation} \label{eq:g-term}
\begin{gathered}
G(\mathrm{x}_{n}, \mathrm{x}_{i}) = \frac{1}{\Vert \mathrm{x}_{n} - \mathrm{x}_{i} \Vert^{2}}
\end{gathered}  
\end{equation} 

$L_{r}(\mathrm{x}_{n}, \mathrm{x}_{i})$ is the attenuated radiance that accounts for the fraction lost during the propagation from the light source until reaching $\mathrm{x}_{i}$ given by the product of the radiance emitted from the light source and the medium transmittance $T_{r}$ between the two points where 

\begin{equation} \label{eq:attenuated-radiance}
\begin{gathered}
L_{r}(\mathrm{x}_{n}, \mathrm{x}_{i}) = \
L_{i}(\mathrm{x}_{n}, \mathrm{x}_{i}) \, T_{r}(\mathrm{x}_{n}, \mathrm{x}_{i})
\end{gathered}  
\end{equation} 

\begin{equation} \label{eq:transmittance}
\begin{gathered}
T_{r}(\mathrm{x}_{n}, \mathrm{x}_{i}) = \
\exp \, \Big(- \displaystyle \int_{\mathrm{x}_{n}}^{\mathrm{x}_{i}} \sigma_t(\mathrm{t}) \, \mathrm{dt} \, \Big) 
\end{gathered}  
\end{equation}

where $\sigma_t = \sigma_a + \sigma_s$ and $ \sigma_a$ are the extinction and absorption coefficients of the medium respectively. 

Equation (\ref{eq:rte-three-point-form}) can be extended to account for the light propagation from $\mathrm{x}_{n}$ on the light source towards $\mathrm{x}_{0}$ on the camera film after multiple scattering events in the medium. Therefore, the radiance recorded by the film due to multiple scattering is computed by the integral of the radiance added from all the other sampled points within the extent of the medium. 
If the light scatters over the path $\overline{x}_{n} = {x}_{0}, {x}_{1}, ..., {x}_{n} \in \Re^{3}$ that consists of $n + 1$ vertices, this radiance can be expressed by the following integral  

 
\begin{equation} \label{eq:rte-three-point-form-multiple-scattering}
\begin{gathered}
L(\mathrm{x}_{0}, \omega)_{\arrowvert{\overline{x}_{n}}} = \, \\
\underbrace{\displaystyle \int_{A}..\, \displaystyle \int_{V}}_{n - 1} \,
L_{\mathrm{r}}(\mathrm{x}_{n}, \mathrm{x}_{n - 1}) \, 
V(\mathrm{x}_{n}, \mathrm{x}_{n - 1}) \, \times  T(\overline{x}_{n}) 
\mathrm{d}V(\mathrm{x}_{2}) \, .. \, \mathrm{d}A(\mathrm{x}_{n})
\end{gathered}  
\end{equation}
 
 
\begin{equation} \label{eq:path-throughput}
\begin{gathered}
T(\overline{x}_{n}) = 
\displaystyle \prod_{i = 1}^{n - 1} \, 
\Big[F(\mathrm{x}_{i + 1}, \mathrm{x}_{i}, \mathrm{x}_{i - 1}) \,
G(\mathrm{x}_{i + 1}, \mathrm{x}_{i}) \hat{V}(\mathrm{x}_{i + 1}, \mathrm{x}_{i}) \Big]
\end{gathered}  
\end{equation}

where $T(\overline{x}_{n})$ is the path throughput that represents the fraction of radiance arriving to $x_{0}$ after $n$ scattering events along the path, and $\hat{V}$ is the attenuated visibility the include the transmittance $Tr$ between the scattering events.

Based on the Monte Carlo fluorescence model presented previously, this rendering integral is extended to handle fluorescence by (1) integrating Equation~(\ref{eq:rte-three-point-form-multiple-scattering}) over all the excitation wavelengths $\lambda_\mathrm{x}$ that illuminate the scene, (2) adding a function to account for the radiance fraction corresponding to the excitation-to-emission event $F_{f}$, and (3) adding a term to indicate the validity of the path. As mentioned before, the path can be only valid if only one fluorescence emission event is sampled. We have added a new binary term in the integral called \emph{fluorescence visibility term} $V_{f}$, whose value is 1 given a valid path and 0 otherwise. The extended integral is shown in Equation~(\ref{eq:rte-three-point-form-multiple-scattering-fluorescence}), where $\lambda_{\mathrm{x}}$ and $\lambda$ are the excitation and emission wavelengths. 

 
\begin{equation} \label{eq:rte-three-point-form-multiple-scattering-fluorescence}
\begin{gathered}
L(\mathrm{x}_{0}, \omega, \lambda)_{\arrowvert{\overline{x}_{n}}} = \, \\
\displaystyle \int_{\lambda_{\mathrm{x}}} \, 
\overbrace{\displaystyle \int_{A}..\, 
    \displaystyle \int_{V}}^{n - 1} \,
L_{\mathrm{r}}(\mathrm{x}_{n}, \mathrm{x}_{n - 1}, \lambda_{\mathrm{x}}) \, 
V(\mathrm{x}_{n}, \mathrm{x}_{n - 1}) \, 
V_{f} \, 
F_{f} (\lambda_{\mathrm{x}}, \lambda) \, \times \\
\displaystyle \prod_{i = 1}^{n - 1} \, 
\Big[F(\mathrm{x}_{i + 1}, \mathrm{x}_{i}, \mathrm{x}_{i - 1}, \lambda) \,
G(\mathrm{x}_{i + 1}, \mathrm{x}_{i}) \,
\hat{V}(\mathrm{x}_{i + 1}, \mathrm{x}_{i}) \Big] \, 
\mathrm{d}V(\mathrm{x}_{2}) \, .. \, 
\mathrm{d}A(\mathrm{x}_{n}) \,
\mathrm{d}\lambda_\mathrm{x}
\end{gathered}  
\end{equation}

The excitation-to-emission function $F_{f}$ depends mainly on the characteristic spectra of fluorophore 

 
\begin{equation} \label{eq:f-function}
\begin{gathered}
F_{f} (\lambda_{\mathrm{x}}, \lambda)= \,
\frac{f_{\mathrm{m}}(\lambda) \, \delta \lambda}{\displaystyle \int_{0}^{\infty} f_{\mathrm{m}}(\lambda') \, \mathrm{d}\lambda'} \cdot f_{\mathrm{x}}(\lambda_{\mathrm{x}})
\end{gathered}  
\end{equation}

where $f_{\mathrm{x}}$  and $f_{\mathrm{m}}$ are the excitation and emission spectra of the dye, and the spectral sampling step $\delta \lambda$ is always set to 1. 

Finally, the radiance reaching $\mathrm{x}_0$ over all the possible paths that exhibit fluorescence emission is computed by 

 
\begin{equation} \label{eq:l-over-all-paths}
\begin{gathered}
L(\mathrm{x}_{0}, \omega, \lambda) = \,
\sum_{n = 1}^{\infty}
L(\mathrm{x}_{0}, \omega, \lambda)_{\arrowvert{\overline{x}_{n}}}
\end{gathered}  
\end{equation}

\subsection{Monte Carlo Estimator}
Using Monte Carlo ray tracing, the total fluorescence in the scene due to multiple scattering is estimated by

\begin{equation} \label{equation:monte-carlo-estimator}
\begin{gathered}
L_{\mathrm{i}}({\mathrm{x}}_{0}, \omega, \lambda) \approx 
\frac{1}{N_{\lambda}} \frac{1}{N} \,
\sum_{\lambda_{\mathrm{x}} = 1}^{N_{\lambda}} \sum_{i = 1}^{N} 
\frac{L_{r}({\mathrm{x}}_{n},\lambda_{\mathrm{x}}) \, V_{i} \, F_{f_{i}}} {p({\mathrm{x}}_{n}) p(\lambda_{{\mathrm{x}}})} 
V_{{f}_{i}} \prod_{j = 1}^{M} \frac{{F_{j}} \, G_{j} \, \hat{V}_{j}}{p(\omega_{j}) p(t_{j})} 
\end{gathered}
\end{equation}

where $p(.)$ is the probability density function for sampling a point $\mathrm{x}_{n}$ on the light source, excitation wavelength $\lambda_{\mathrm{x}}$, an elastic scattering event with direction $\omega_{j}$ and distance $t_{j}$. $N$ and $N_{\lambda}$ are the numbers of paths and spectral samples.


\section{Results: Rendering Fluorescent Solutions}
Our fluorescence path tracer is employed to render fluorescent solutions containing multiple fluorophore dyes at different concentrations. The dyes have been selected from the Alexa Fluor family, which has significant applications in biology and spectroscopy. Seven dyes have been chosen carefully to cover the visible spectrum including Alexa Fluor 350 (blue), 405 (violet), 488 (green), 546 (yellow) , 568 (orange), 610 (red) and 633 (far red) ~\cite{johnson2010the}. The dyes are dissolved in a highly scattering solvent with a scattering coefficient that is 100 times greater than that of pure water. The spectroscopic properties of the fluorophores are obtained from an online database available from Thermo Fisher Scientific~\cite{spectraviewer2016}. 

\begin{figure*}[!ht]
    \centering
    \mbox{} \hfill
    \includegraphics[width=\textwidth]{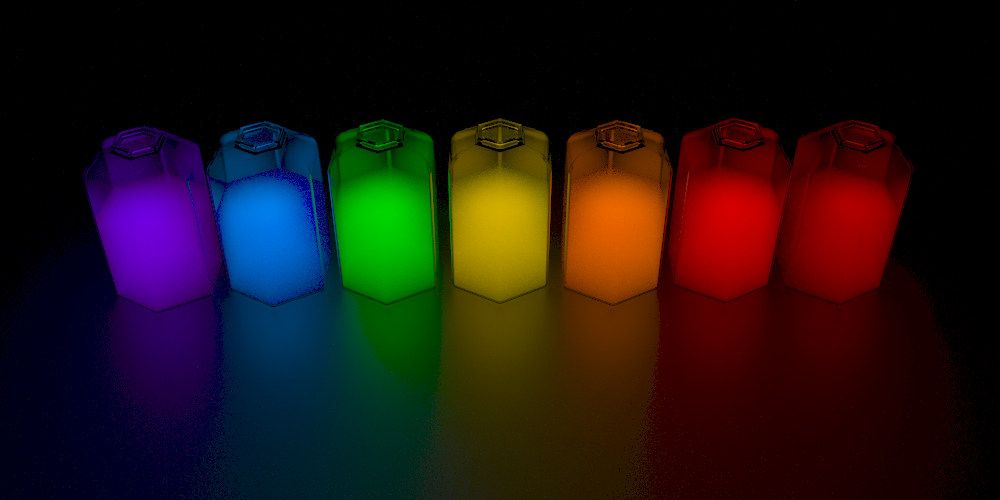}
    \hfill \mbox{}
    \caption{\label{fig:fluorescent-vials-spectrum}
        The fluorescent vials are filled with different fluorophores from the Alexa Fluor family dissolved in highly scattering solution. The scene is illuminated with multiple wavelengths corresponding to the maximum excitation of each dye to achieve maximum emission.  
    }
\end{figure*}

Figure~(\ref{fig:teaser-fluorescent-spheres}) shows a glossy Cornell box containing group of transparent fluorescent spheres (beads) filled with the selected Alexa Fluor solutions. The spheres are arranged from right to left according to the maximum emission wavelength of each fluorophore. 
All the solutions have the same fluorophore concentration. 
The scene is excited with monochromatic illumination from different diffuse light sources the emit at the maximum excitation wavelength of each respective fluorophore. In Figure~(\ref{fig:fluorescent-vials-spectrum}), we replace the transparent spheres with multiple glass vials containing the same set of fluorescent solutions.  

\begin{figure}[!h]
    \centering
    \mbox{} \hfill
    \includegraphics[width=0.5\textwidth]{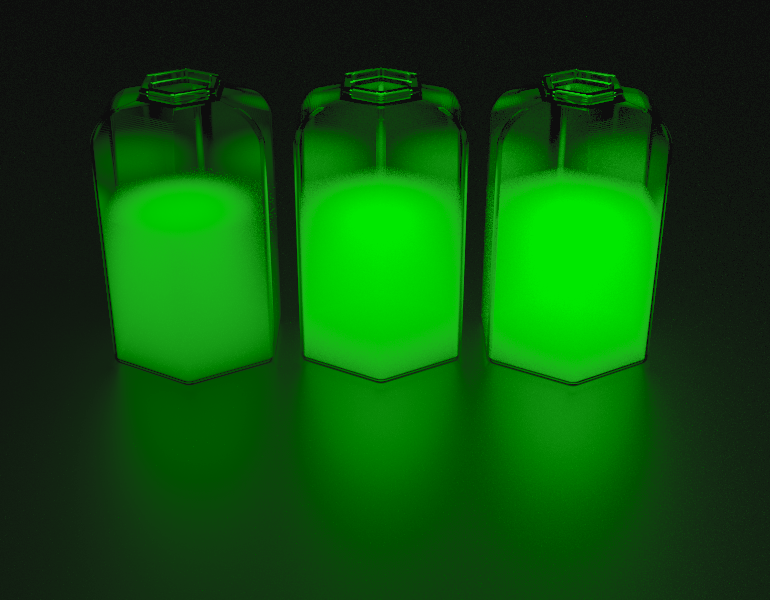}
    \hfill \mbox{}
    \caption{\label{fig:fluorescent-vials-concentrations}
        Fluorescent vials containing Alexa Fluor 488 solutions having three different dye concentrations: 0.1, 1 and 10 grams per liter. 
    }
\end{figure}

In Figure~(\ref{fig:fluorescent-vials-concentrations}), we demonstrate the results of varying the fluorophore concentration under fixed illumination conditions and for the same dye. The vials contain Alexa Fluor 488 solutions concentrated at 0.1, 1 and 10 grams per liter. 
The vials are illuminated at 495 nm, corresponding to the maxim excitation wavelength of the dye. 
According to Beer-Lambert law and the Brightness equation, increasing dye concentration in the solution is accompanied with more fluorescence emission. This is quite obvious from the glossy reflection of the fluorescent light in front of each vial.
All the images have been rendered at 1 nm of spectral sampling (500 samples) and 512 $\times$ 512 samples per pixel. 

\section{Model Validation}
To evaluate the spectral performance of the presented fluorescence model, we validated the emission \acs{SPD} profiles computed from the renderings against the emission spectra of pure fluorophores that were recorded in real spectroscopic experiments~\cite{spectraviewer2016}. The \acs{SPD}s corresponding to all the pixels in the image are obtained before their conversion to RGB colors. The final SPD that represents the spectral response of the entire image is normalized by the total number of illuminated pixels. This \acs{SPD} reflects the emission recorded by the camera from all the fluorescent objects in the scene. 

\begin{figure*}[!ht]
    \centering
    \mbox{} \hfill
    \includegraphics[width=\textwidth]{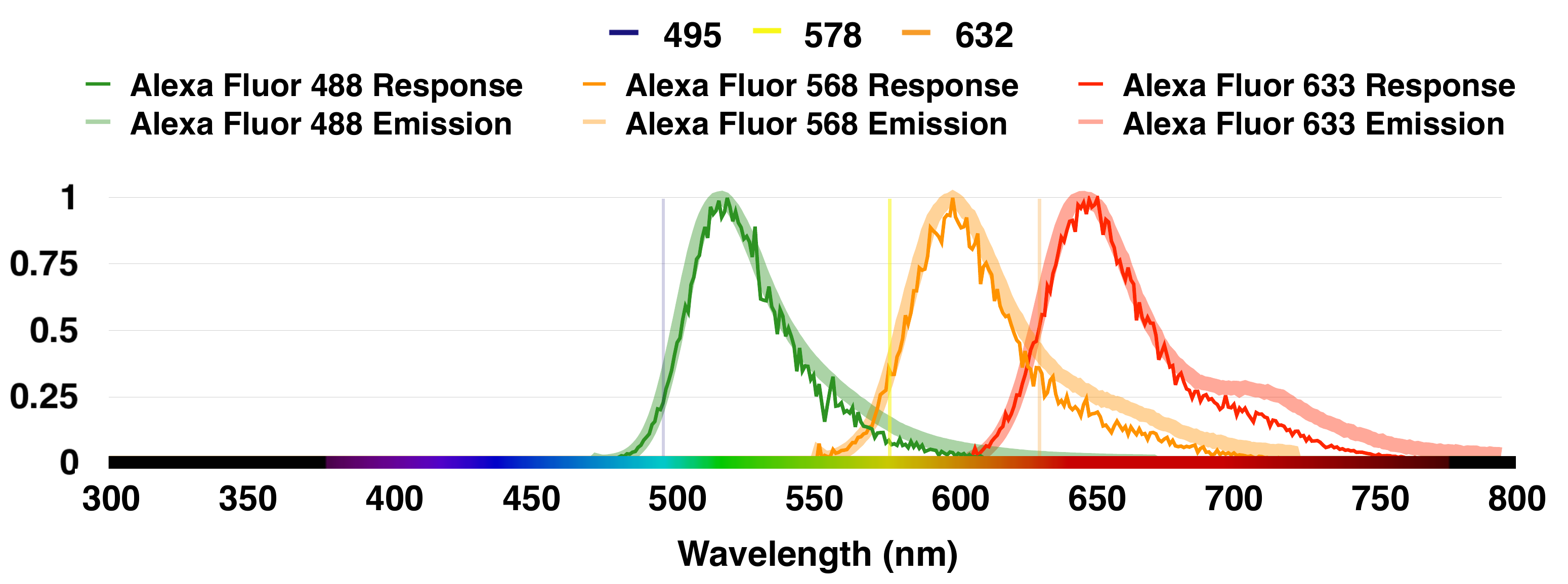}
    \hfill \mbox{}
    \caption{\label{fig:validation-maximum-response}
        Normalized emission \acs{SPD}s computed from rendering three fluorescent beads filled with Alexa Fluro 488, 568 and 633 solutions. The profiles of these \acs{SPD}s are compared against the normalized emission spectra of the pure dyes. The \acs{SPD}s are obtained when the beads are illuminated at the maximum excitation wavelengths of each respective dye. 
    }
\end{figure*}

The validation of the model is addressed relying on two complementary tests. The first one compares the profiles of the \acs{SPD}s computed from the images with respect to the intrinsic emission spectra of the pure fluorophores. The experimental spectra of pure fluorophores are recorded from fluorescent solutions with negligible scattering coefficients. 
As a consequence of the wavelength dependence of the optical properties of the solvents, the emission spectra detected from highly scattering solutions cannot not perfectly match those obtained from pure fluorophores. Though, the normalized spectra recorded from those solutions should have similar profiles~\cite{liu2012experimental, valeur2012molecular}. We have designed three test scenes containing spheres filled with Alexa Fluor 488, 568 and 632 solutions to validate this aspect.
The spheres were illuminated at 495, 578, and 632 nm corresponding to the maximum excitation wavelength of each respective dye. Figure~(\ref{fig:validation-maximum-response}) shows the \acs{SPD}s obtained from rendering each solution individually in comparison with the real emission spectra of the pure dyes. 

\begin{figure*}[htpb]
    \centering
    \mbox{} \hfill
    \includegraphics[width=\textwidth]{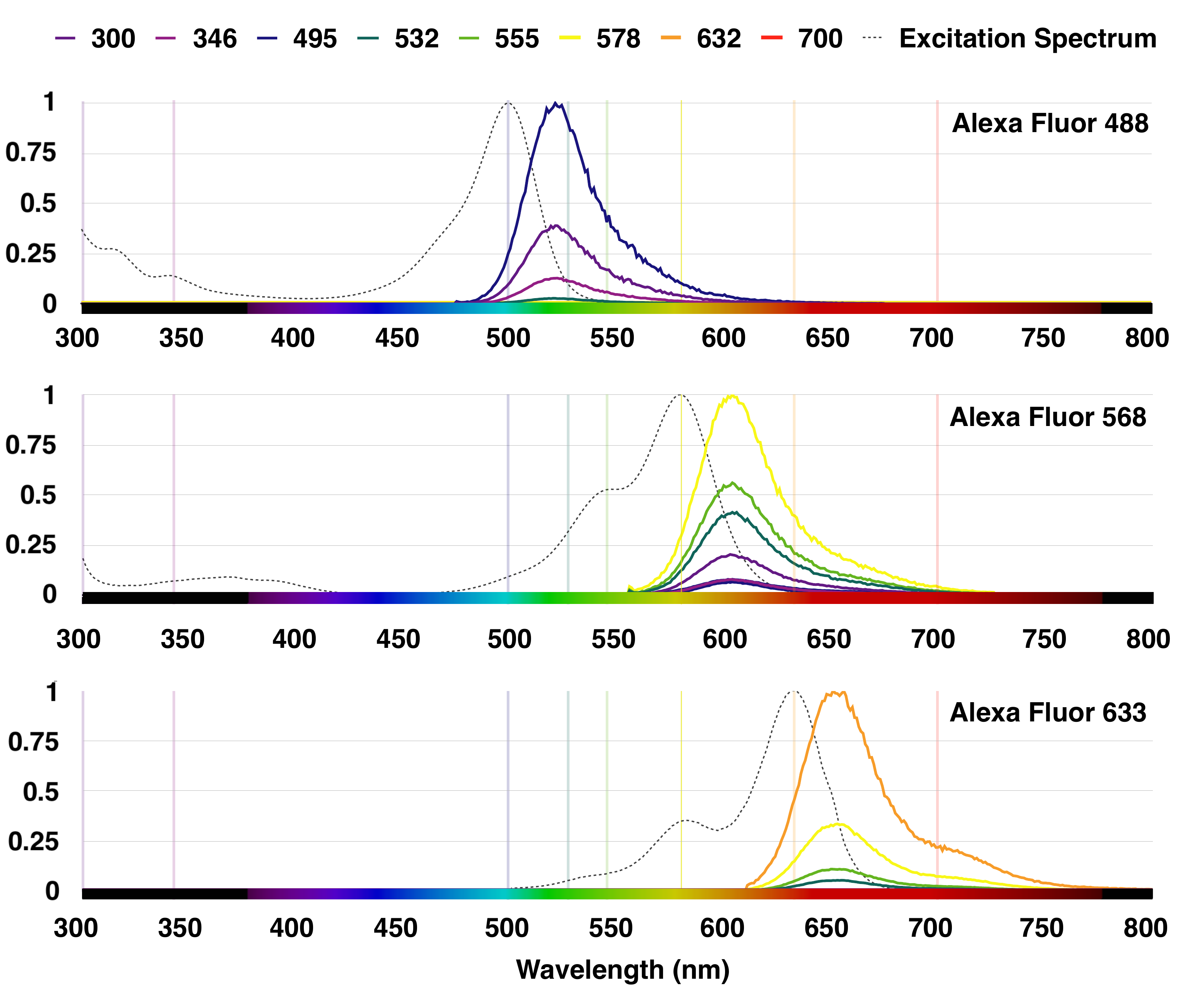}
    \hfill \mbox{}
    \caption{\label{fig:validation-different-wavelengths}
        Relative emission \acs{SPD}s computed from rendering three fluorescent beads filled with Alexa Fluro 488, 568 and 633 solutions. The beads are illuminated at multiple excitation wavelengths between 300 and 700 nm. The profiles are normalized with respect to the emission SPD measured at maximum emission wavelength for each respective dye. 
    }
\end{figure*}

The other validation test measures the accuracy of the model when the fluorescent solutions are illuminated with different excitation wavelengths. As shown in Figure~(\ref{fig:fluorescence-physics}), varying the excitation wavelength does not change the profile of the emission spectrum, but scales it 
proportional to the amplitude of the excitation spectrum at the excitation wavelength. The same scene used in the first test are illuminated with multiple excitation wavelengths between 300 and 700 nm. The results of this test are shown in Figure~(\ref{fig:validation-different-wavelengths}). 

\section{Conclusions}
We presented a physically-accurate Monte Carlo model for rendering fluorescence. The model accounts for the intrinsic characteristics of fluorescent compounds including their excitation and emission spectra, quantum yield and molar absorptivity in addition to their concentration in the medium. 
The formalism of the standard path tracing algorithm is adapted to treat highly scattering participating media in the presence of fluorescent particles. 
This adaptation results in unbiased estimator that is guaranteed to converge to the exact solution of the full \acs{RTE}.
The results of our rendering algorithm are demonstrated to complex scenes containing different fluorescent solutions having multiple concentrations.
The model is quantitatively validated against the emission spectra of realistic fluorophores. Our method is expected to be applied in computational photography and in silico neurobiology. 

\newpage

\begin{thebibliography}{\uppercase{WGRK{\etalchar{*}}97}}
    
    \bibitem[ABE{\etalchar{*}}15a]{abdellah2015computational}
    \textsc{Abdellah M., Bilgili A., Eilemann S., Markram H., Sch{\"u}rmann F.}:
    \newblock A computational model of light-sheet fluorescence microscopy using
    physically-based rendering.
    \newblock In \emph{Eurographics 2015} (2015), The European Association for
    Computer Graphics (Eurographics), p.~2.
    
    \bibitem[ABE{\etalchar{*}}15b]{abdellah2015physically}
    \textsc{Abdellah M., Bilgili A., Eilemann S., Markram H., Sch{\"u}rmann F.}:
    \newblock Physically-based in silico light sheet microscopy for visualizing
    fluorescent brain models.
    \newblock \emph{BMC bioinformatics 16}, Suppl 11 (2015), S8.
    
    \bibitem[CPP{\etalchar{*}}05]{cerezo2005survey}
    \textsc{Cerezo E., P{\'e}rez F., Pueyo X., Seron F.~J., Sillion F.~X.}:
    \newblock A survey on participating media rendering techniques.
    \newblock \emph{The Visual Computer 21}, 5 (2005), 303--328.
    
    \bibitem[CS03]{cerezo2003inelastic}
    \textsc{Cerezo E., Seron F.}:
    \newblock Inelastic scattering in participating media. application to the
    ocean.
    \newblock In \emph{Proceedings of the Annual Conference of the European
        Association for Computer Graphics, Eurographics 2003} (2003).
    
    \bibitem[CS04]{cerezo2004rendering}
    \textsc{Cerezo B.~E., Seron F.~J.}:
    \newblock Rendering natural waters taking fluorescence into account.
    \newblock \emph{Computer Animation and Virtual Worlds 15}, 5 (2004), 471--484.
    
    \bibitem[Gla95]{glassner1995principles}
    \textsc{Glassner A.~S.}:
    \newblock \emph{Principles of Digital Image Synthesis}.
    \newblock Morgan Kaufmann, 1995.
    
    \bibitem[GSMA04]{gutierrez2004inelastic}
    \textsc{Gutierrez D., Seron F.~J., Mu{\~n}oz A., Anson O.}:
    \newblock Inelastic scattering in participating media using curved photon
    mapping.
    \newblock In \emph{ACM SIGGRAPH 2004 Sketches} (2004), ACM, p.~76.
    
    \bibitem[GSMA08]{gutierrez2008visualizing}
    \textsc{Gutierrez D., Seron F.~J., Munoz A., Anson O.}:
    \newblock Visualizing underwater ocean optics.
    \newblock In \emph{Computer Graphics Forum} (2008), vol.~27, Wiley Online
    Library, pp.~547--556.
    
    \bibitem[HHA{\etalchar{*}}10]{hullin2010acquisition}
    \textsc{Hullin M.~B., Hanika J., Ajdin B., Seidel H.-P., Kautz J., Lensch H.}:
    \newblock Acquisition and analysis of bispectral bidirectional reflectance and
    reradiation distribution functions.
    \newblock \emph{ACM Transactions on Graphics (TOG) 29}, 4 (2010), 97.
    
    \bibitem[Jak10]{mitsuba2010}
    \textsc{Jakob W.}:
    \newblock Mitsuba, physically-based rendering, 2010.
    
    \bibitem[JC98]{jensen1998efficient}
    \textsc{Jensen H.~W., Christensen P.~H.}:
    \newblock Efficient simulation of light transport in scenes with participating
    media using photon maps.
    \newblock In \emph{Proceedings of the 25th annual conference on Computer
        graphics and interactive techniques} (1998), ACM, pp.~311--320.
    
    \bibitem[JDL{\etalchar{*}}14]{jiang2014evaluation}
    \textsc{Jiang X., Deng Y., Luo Z., Wang K., Lian L., Yang X., Meglinski I., Luo
        Q.}:
    \newblock Evaluation of path-history-based fluorescence monte carlo method for
    photon migration in heterogeneous media.
    \newblock \emph{Optics express 22}, 26 (2014), 31948--31965.
    
    \bibitem[Joh10]{johnson2010the}
    \textsc{Johnson I.}:
    \newblock \emph{The Molecular Probes Handbook: A Guide to Fluorescent Probes
        and Labeling Technologies, 11$^th$ Edition}.
    \newblock 2010.
    \newblock Life Technologies Corporation.
    
    \bibitem[JS10]{johnson2010molecular}
    \textsc{Johnson I., Spence M.}:
    \newblock The molecular probes handbook.
    \newblock \emph{Life Technologies Corporation} (2010).
    
    \bibitem[Kaj86]{kajiya1986rendering}
    \textsc{Kajiya J.~T.}:
    \newblock The rendering equation.
    \newblock In \emph{ACM Siggraph Computer Graphics} (1986), vol.~20, ACM,
    pp.~143--150.
    
    \bibitem[LDW{\etalchar{*}}15]{luo2015decoupled}
    \textsc{Luo Z., Deng Y., Wang K., Lian L., Yang X., Luo Q.}:
    \newblock Decoupled fluorescence monte carlo model for direct computation of
    fluorescence in turbid media.
    \newblock \emph{Journal of biomedical optics 20}, 2 (2015), 025002--025002.
    
    \bibitem[LRV{\etalchar{*}}12]{liu2012experimental}
    \textsc{Liu C., Rajaram N., Vishwanath K., Jiang T., Palmer G.~M., Ramanujam
        N.}:
    \newblock Experimental validation of an inverse fluorescence monte carlo model
    to extract concentrations of metabolically relevant fluorophores from turbid
    phantoms and a murine tumor model.
    \newblock \emph{Journal of biomedical optics 17}, 7 (2012), 0780031--07800315.
    
    \bibitem[LS13]{lam2013spectral}
    \textsc{Lam A., Sato I.}:
    \newblock Spectral modeling and relighting of reflective-fluorescent scenes.
    \newblock In \emph{Proceedings of the IEEE Conference on Computer Vision and
        Pattern Recognition} (2013), pp.~1452--1459.
    
    \bibitem[lux13]{luxrender2013}
    Luxrender, gpl physically-based renderer, 2013.
    
    \bibitem[MSHD15]{meng2015physically}
    \textsc{Meng J., Simon F., Hanika J., Dachsbacher C.}:
    \newblock Physically meaningful rendering using tristimulus colours.
    \newblock In \emph{Computer Graphics Forum} (2015), vol.~34, Wiley Online
    Library, pp.~31--40.
    
    \bibitem[PH10]{pharr2004physically}
    \textsc{Pharr M., Humphreys G.}:
    \newblock \emph{Physically Based Rendering, Second Edition: From Theory To
        Implementation}, 2$^{nd}$~ed.
    \newblock Morgan Kaufmann Publishers Inc., San Francisco, CA, USA, 2010.
    
    \bibitem[PH12]{pbrt2012}
    \textsc{Pharr M., Humphreys G.}:
    \newblock Physically-based rendering. from theory to implementation, 2012.
    
    \bibitem[PKK00]{pauly2000metropolis}
    \textsc{Pauly M., Kollig T., Keller A.}:
    \newblock Metropolis light transport for participating media.
    \newblock In \emph{Rendering Techniques 2000}. Springer, 2000, pp.~11--22.
    
    \bibitem[PM93]{pattanaik1993computation}
    \textsc{Pattanaik S.~N., Mudur S.~P.}:
    \newblock Computation of global illumination in a participating medium by monte
    carlo simulation.
    \newblock \emph{The Journal of Visualization and Computer Animation 4}, 3
    (1993), 133--152.
    
    \bibitem[RSK08]{raab2008unbiased}
    \textsc{Raab M., Seibert D., Keller A.}:
    \newblock Unbiased global illumination with participating media.
    \newblock In \emph{Monte Carlo and Quasi-Monte Carlo Methods 2006}. Springer,
    2008, pp.~591--605.
    
    \bibitem[Sci16]{spectraviewer2016}
    \textsc{Scientific T.~F.}:
    \newblock Fluorescence spectra viewer, 2016.
    
    \bibitem[SG95]{glassner1995model}
    \textsc{S.~Glassner A.}:
    \newblock A model for fluorescence and phosphorescence.
    \newblock In \emph{Photorealistic Rendering Techniques}. Springer, 1995,
    pp.~60--70.
    
    \bibitem[SLW08]{sharpe2008silico}
    \textsc{Sharpe J., Lumsden C.~J., Woolridge N.}:
    \newblock \emph{In silico: 3D animation and simulation of cell biology with
        Maya and MEL}.
    \newblock Morgan Kaufmann, 2008.
    
    \bibitem[SPEAE03]{swartling2003accelerated}
    \textsc{Swartling J., Pifferi A., Enejder A.~M., Andersson-Engels S.}:
    \newblock Accelerated monte carlo models to simulate fluorescence spectra from
    layered tissues.
    \newblock \emph{JOSA A 20}, 4 (2003), 714--727.
    
    \bibitem[Tec10]{render2010maxwell}
    \textsc{Technologies N.~L.}:
    \newblock Maxwell render 2.5 user manual, 2010.
    
    \bibitem[VBS12]{valeur2012molecular}
    \textsc{Valeur B., Berberan-Santos M.~N.}:
    \newblock \emph{Molecular Fluorescence: Principles and Applications}.
    \newblock John Wiley \& Sons, 2012.
    
    \bibitem[WGRK{\etalchar{*}}97]{welch1997propagation}
    \textsc{Welch A., Gardner C., Richards-Kortum R., Chan E., Criswell G., Pfefer
        J., Warren S.}:
    \newblock Propagation of fluorescent light.
    \newblock \emph{Lasers in surgery and medicine 21}, 2 (1997), 166--178.
    
    \bibitem[WK90]{wolff1990ray}
    \textsc{Wolff L.~B., Kurlander D.~J.}:
    \newblock Ray tracing with polarization parameters.
    \newblock \emph{Computer Graphics and Applications, IEEE 10}, 6 (1990), 44--55.
    
    \bibitem[WND{\etalchar{*}}14]{wilkie2014hero}
    \textsc{Wilkie A., Nawaz S., Droske M., Weidlich A., Hanika J.}:
    \newblock Hero wavelength spectral sampling.
    \newblock In \emph{Computer Graphics Forum} (2014), vol.~33, Wiley Online
    Library, pp.~123--131.
    
    \bibitem[WTP01]{wilkie2001combined}
    \textsc{Wilkie A., Tobler R.~F., Purgathofer W.}:
    \newblock \emph{Combined rendering of polarization and fluorescence effects}.
    \newblock Springer, 2001.
    
    \bibitem[WW11]{wilkie2011physically}
    \textsc{Wilkie A., Weidlich A.}:
    \newblock A physically plausible model for light emission from glowing solid
    objects.
    \newblock In \emph{Computer Graphics Forum} (2011), vol.~30, Wiley Online
    Library, pp.~1269--1276.
    
    \bibitem[WWLP06]{wilkie2006reflectance}
    \textsc{Wilkie A., Weidlich A., Larboulette C., Purgathofer W.}:
    \newblock A reflectance model for diffuse fluorescent surfaces.
    \newblock In \emph{Proceedings of the 4th international conference on Computer
        graphics and interactive techniques in Australasia and Southeast Asia}
    (2006), ACM, pp.~321--331.
    
\end{thebibliography}

\section*{Acronyms}
\begin{acronym}
    \acro{3D}{Three-dimensional}
    \acro{BBP}{Blue Brain Project}
    \acro{BRDF}{Bidirectional Reflectance Distribution Function}
    \acro{BRRDF}{Bidirectional Reflectance \& Reradiation Distribution Function}
    \acro{CCD}{Charged Coupled Device}
    \acro{CFP}{Cyan Fluorescent Protein}
    \acro{CPU}{Central Processing Unit}
    \acro{CG}{Computer Graphics}
    \acro{DIC}{Differential Interference Contrast}
    \acro{FBE}{Fluorescence Brightness Equation}
    \acro{FRTE}{Full Radiative Transfer Equation}
    \acro{FOV}{Field of View}
    \acro{GFP}{Green Fluorescent Protein}
    \acro{GPU}{Graphics Processing Unit}
    \acro{HBP}{Human Brain Project}
    \acro{LSFM}{Light Sheet Fluorescence Microscopy}
    \acro{MRI}{Magnetic Resonance Imaging}
    \acro{NA}{Numeric Aperture}
    \acro{PBR}{Physically-based Rendering}
    \acro{PDF}{Probability Density Function}
    \acro{PBRT}{Physically-based Rendering Toolkit}
    \acro{PCM}{Phase Contrast Microscopy}
    \acro{PSF}{Point Spread Function}
    \acro{RFP}{Red Fluorescent Protein}
    \acro{RTE}{Radiative Transfer Equation}
    \acro{SPD}{Spectral Power Distribution}
    \acro{SFP}{Simulated Fluorescence Process}
    \acro{YFP}{Yellow Fluorescent Protein}
    \acro{VSDI}{Voltage Sensitive Dye Imaging}
\end{acronym}

\newcommand{\etalchar}[1]{$^{#1}$}

\end{document}